\documentstyle[eqsecnum,aps,epsfig]{revtex}

\newcommand{\beq}{\begin{equation}}
\newcommand{\eeq}{\end{equation}}

\begin{document}

\title{Raman Spectroscopic Investigation of H$_2$, HD, and D$_2$ 
Physisorption on Ropes of Single-Walled, Carbon Nanotubes}

\author{Keith~A.~Williams, Bhabendra~K.~Pradhan, and Peter~C.~Eklund\footnote{To whom 
correspondence should be addressed; email: {\it pce3@psu.edu} }\\ 
Milen~K.~Kostov and Milton~W.~Cole}

\address{Physics Department, The Pennsylvania State University,\\ 
104 Davey Laboratory, University Park, PA 16802}

\date{\today}
\maketitle
\begin{abstract}

We have observed the S- and Q-branch Raman spectra of H$_2$, HD, and
D$_2$ adsorbed at 85~K and pressures up to 8~atm on single-walled,
carbon nanotubes (SWNT).  Comparative data for H$_2$ on graphite and
C$_{60}$ were also collected.  For each adsorbate, we observed a small
shift in the Q-branch frequencies relative to the gas-phase values.  To
aid in interpreting this result, we constructed an H$_2$-surface
potential, including van der Waals and electrostatic terms.  Computed
shifts based on this potential are in good agreement with our data.

\end{abstract}

\vspace{0.3 in}
%----------------------------------------------------------INTRO

Single-walled, carbon nanotubes (SWNT) are nanoporous, all-surface
macromolecules which may be ideal storage media for
H$_2$\cite{dresselhaus,williams}.  In addition, H$_2$ has been
predicted to exhibit novel, quasi-1D phases in SWNT
ropes\cite{calbi,cole}.  Desorption experiments suggest the presence of
a high-energy binding site for H$_2$ in purified and sonicated SWNT
material, which has been proposed to be indicative of charge
transfer\cite{dillonheben,heben}.  On the other hand, transport data
for SWNT suggest that H$_2$ physisorbs\cite{adu}.  An improved
understanding of the adsorption of H$_2$ on SWNT is the motivation for
the work described here.  We have collected the S- and Q-branch Raman
spectra of H$_2$, HD, and D$_2$ adsorbed on SWNT at 85~K and at
pressures up to 8~atm.  The magnitude of the shifts in the Q-branch
frequencies of these species when adsorbed on SWNT is a measure of the
strength of the adsorption potential.  For physisorption, the shifts
should be small, while in the event of charge transfer, the shifts
should be large: the change of the H$_2$ stretching frequency with
the molecular charge is $\sim$2000~cm$^{-1}$ per
electron\cite{huber}.  The approach taken in this work is similar to
that described in Ref.\cite{conner}, except that we have employed Raman
rather than IR spectroscopy.

Using a two-step, oxidative and acid reflux technique, we obtained SWNT
samples containing minimal amorphous and multishell carbon
($<$15~wt-\%) and residual metal catalyst ($<$10~wt-\%,
$<$2~atom-\%), as determined by temperature programmed oxidation.
This material was annealed in a vacuum of better than 10$^{-7}$~Torr at
1200~K for 24~hours; such high-temperature ``degassing'' was recently
shown to remove the carboxylic and related functional groups occluding
the SWNT ends and side walls\cite{yates}.  The resulting material was
pressed against indium foil and inserted in an optical pressure cell, which
was placed in a cryostat (Janis Corp.  Model VPF 700, 77-700~K) with a
gas line to the cell. The cell was valved to a diffusion pump and to
the cylinders of H$_2$ (ultra-high purity, 99.999\%, MG~Industries), HD
(97.4 mol-\%, with 1.4\% H$_2$ and 1.2\% D$_2$, Isotec~Inc.), and D$_2$
(ultra-high purity, 99.999\%, MG~Industries).  In separate experiments,
samples of C$_{60}$ powder (Alfa~Aesar, 99.9\%, packed under Ar) and
freshly cleaved, pyrolytic graphite (Alfa~Aesar) were loaded into our
cell.

Raman spectra were collected in the backscattering geometry, with
$\lambda_{air}$=514.53~nm excitation provided by a mixed-gas, Ar/Kr-ion
laser (Coherent Inc., Innova Spectrum).  The incident beam was focused
through the windows of the cryostat and pressure cell onto the sample
surface.  Backscattered Rayleigh light was rejected by a holographic
``supernotch'' filter (Kaiser, Inc.), and Raman light was focused
through a 25~$\mu$m slit into a single-grating monochromator
(Instruments S.A.  HR460, grating:  1800~grooves/mm) equipped with a
CCD. To correct our spectra for slight, instrumental nonlinearity and
to calibrate the spectra, one Hg line and several pairs of Ne lines
close to the Q-branch lines of H$_2$, D$_2$, and HD were used. The Ar,
Hg and Ne frequencies were taken from Refs.\cite{norlen} and
\cite{burns}.

%-----------------------------------------------------------RESULTS

For all data reported here, the sample temperature was 85$\pm$5~K, and
the pressure of the adsorbate gas on the sample was increased from
vacuum to as high as 8~atm, in the case of H$_2$ and D$_2$, and to
5~atm in the case of HD.  Typical S- and Q-branch spectra at 8~atm are
shown in Figs.1 and 2; we use the notation $S(J)$ or $Q(J)$ to denote
transitions, in which $J$ is the initial state.  The frequencies,
$\nu_{ads}$, of the observed S- and Q-branch transitions are given in
Table~1; these values were determined by fitting Voigt-profile
functions to the spectra using Huml\'\i\v{c}ek's
algorithm\cite{humlicek}.  Also provided in Table-1 are the
frequencies, $\nu_{free}$, of these transitions for free molecules,
from Ref.\cite{huber}.

Our S-branch data suggest freely rotating molecules in the gas phase,
and are in good agreement with recent inelastic neutron scattering
data\cite{brown}.  Small, downshifted shoulders are evident in the
S-branch lines, but our experimental resolution is insufficient to
quantify these effects.  Our Q-branch data for HD and D$_2$ contain
prominent bands which are slightly upshifted from the free-molecule
frequencies (see Table~1), but, due to the small splitting of the J=0
and J=1 bands, the peak substructure could not be resolved.  Our data
for H$_2$ in SWNT are considerably more illuminating, and are shown in
expanded form in Fig.3 at two pressures, along with comparative data
for C$_{60}$ and highly-oriented pyrolytic graphite (HOPG) collected at
85$\pm$5~K and 4~atm.  The best fits (solid curves) to the data in
Fig.3 were computed using a sum of Voigt components (dashed curves).

The surface area of HOPG is extremely small, so the signal strength
from adsorbed molecules should be negligible.  Therefore, we assign the
bands observed at 4161.3 and 4155.4~cm$^{-1}$ to gas-phase H$_2$ near
the surface; these frequencies are in reasonable agreement with the
those reported by May, $et~al$\cite{may}.  The gas-phase components
should also be present in the data on C$_{60}$ and SWNT, and so our
analysis of these data proceeded by inserting components of frequency
and width equal to those observed on HOPG, and then introducing
separate Voigt components to describe the additional bands.  Consistent
with this approach, we noted an approximately  twofold increase in the
relative intensities of the bands attributed to gas-phase species, as
the pressure was increased from 4 to 8~atm.  In our analysis of the
data on SWNT at 8~atm, the positions of the bands attributed to the gas
phase were allowed to vary, and the frequencies were found to be
$\sim$0.1~cm$^{-1}$ lower, again consistent with the data of May,
$et~al$\cite{may}.

In addition to the gas-phase components, it is clear that the data for
C$_{60}$ and SWNT contain additional bands which are lacking in the
data on HOPG.  We ascribe these components to adsorption in the
interstices or internal pores in the respective lattices.  Neutron
powder diffraction work by FitzGerald, $et~al.$, has indicated the
existence of an adsorbed phase in the octahedral sites of
$fcc$-C$_{60}$\cite{fitzgerald}; presumably, these sites are partially
filled under the conditions of our study and account for the
lower-frequency component present in our Q-branch data for H$_2$ on
C$_{60}$.

A reduction in the number of fitting parameters was achieved
by setting equal the linewidth parameters of equally-spaced pairs of
Q(0) and Q(1) peaks.  All parameters were then refined until $\chi^2$
was minimized and the frequencies were obtained.  Our confidence in the
number of Voigt components introduced for each data set is affirmed by
the relative intensities between ``pairs'' of Voigt components:
I$_{Q(1)}$/I$_{Q(0)}$=2.2--2.5, in good agreement with the equilibrium
ortho-para ratio at 85~K.  Also, the Q(1)-Q(0) spacing we
find between these pairs (5.8--6.0~cm$^{-1}$) is consistent with the
value ($\sim$5.97~cm$^{-1}$) reported by May, $et~al$, under similar
conditions\cite{may}.

Based on shifts observed in the tangential Raman bands of SWNT material
previously dosed with H$_2$, partial electron transfer from SWNT to the
H$_2$ adsorbate has been proposed\cite{heben}.  Our experiment is
uniquely suited to looking for such effects, given the extreme
sensitivity of the H$_2$ stretching frequency on the molecular charge
state.  Contrary to reports on certain zeolites and
oxides\cite{conner}, no strongly shifted Q-branch lines were observed
on SWNT.  This would seem to rule out significant charge transfer under
our experimental conditions.

%-----------------------------------------------------------THEORY

To aid in interpreting the small shifts found in our Q-branch data, we
have constructed an interaction potential for H$_2$ with graphene (a
single sheet of graphite) and estimated the frequency shifts in two
types of adsorption sites. For the case of H$_2$ adsorbed on graphene,
the potential is a sum of C-H van der Waals interactions, $U_{LJ}$, and
the electrostatic interaction, $U_{el}$, of the H$_2$ static multipole
moments with the static screening charges induced on the graphene
surface\cite{bruch}.  Because the low-frequency dielectric response of
graphene is metallic\cite{phil}, the screening charges may be
represented by full image charges displaced from the graphene sheet by
a distance $z_0$.  The electrostatic interaction can be expressed in
terms of the H$_2$ quadrupole ($\Theta$) and hexadecapole ($\Phi$)
moments\cite{bruch}:

\beq
\Phi_{el}(z_0, \theta) = -\frac{3k\Theta^2}{8(2z_I)^5}[3\cos^4\theta + 
	2\cos^2\theta +3] - \frac{15k\Theta \Phi}{16(2z_I)^7}[7\cos^6\theta + 
	5\cos^4\theta + 9\cos^2\theta - 5] \, ,
\eeq

\noindent where $z_c$ is the perpendicular distance from the molecular
center to the surface atoms, $z_I=z_c-z_0$ is the location of the image
with respect to the graphene surface, $\theta$ is the angle of the
molecular axis with respect to the surface normal, and $k$ is the usual
electrostatic constant. In view of the small surface corrugation, the
holding potential is $V(z_c,\cos\theta) = U_{el} + U_{LJ}(z_1) +
U_{LJ}(z_2)$, where the $z_i$ are the perpendicular distances of the H
atoms relative to the surface, and the $U_{LJ}(z_i)$ are the
corresponding, pairwise van der Waals interactions.  The $C-H$
interactions are assumed to be of Lennard-Jones ($12$-$6$) form, and
for simplicity, we smear out the C atoms along the surface.  In this
approximation,

\beq
\label{lj}
U_{LJ}(z) =
2\pi\vartheta\epsilon_{C-H}\sigma_{C-H}^{2}\Big[\frac{2}{5}
	(\frac{\sigma_{C-H}}{z})^{10}
	- (\frac{\sigma_{C-H}}{z})^{4}\Big] \, ,
\eeq

\noindent where $\vartheta$ = $0.38$~\AA$^{-2}$ is the surface density
of the C atoms.  We adopted the values $\epsilon_{C-H}=2.26$~meV , 
$\sigma_{C-H}=2.76$~\AA\cite{jan}.

The least well-known parameter in our potential is $z_0$\cite{bruch2}.
To obtain this parameter, we computed the minimum of the C-H$_2$
potential at the equilibrium distance $z_{eq}$, where H$_2$ is
preferentially oriented flat against the surface.  In the case of a
graphene sheet, the isotropic C-H$_2$ holding potential has the same
functional form as Eqn.(\ref{lj}), with a minimum at
$\displaystyle{z_{eq} = \sigma_{C-H_2}}$.  Using
$\epsilon_{C-H_2}$=42.8~K  and $\sigma_{C-H_2}$=2.97\AA\cite{wang}, the
resulting well depth is 46.6~meV. The values for the H$_2$ moments,
taken from Ref.\cite{mar}, are $\Theta = 0.48917$~a.u.; $\Phi =
0.230063$~a.u.. These values yield $z_0 = 1.36$~\AA, a value consistent
with previous {\it ab initio} and empirical estimates for this
system\cite{kim}.

For the study of H$_2$ in the ICs, we adapted a potential,
$U_{LJ}^{\ast}$, already developed for a single nanotube\cite{stan1}.
The dispersion part of this potential ignores many-body effects,
considers the C atoms to be smeared on the surface, and is isotropic.
When averaged over the azimuthal and longitudinal coordinates, this
potential takes the form\cite{stan1}:

\beq
 U_{LJ}^{\ast}( z ) =
 3\pi\vartheta\epsilon_{C-H}\sigma_{C-H}^{2}\Big[\frac{21}{32}
	\Big(\frac{\sigma_{C-H}}{R}\Big)^{10}
 f_{11}(x)M_{11}(x) - \Big(\frac{\sigma_{C-H}}{R}\Big)^{4} f_5(x)
 M_5(x)\Big] \, ,
\eeq

\noindent in which $z$ is the distance from the axis of a nanotube with
nuclear radius $R$; we define $x=R/z$ and $f_l(x)=(R/z)^l$ , with the
$l$ being positive integers.  The $M_l$ are integrals defined in
Ref.\cite{stan1}.  The potential for an IC is obtained by summing
$U_{LJ}^{\ast}$ over an assembly of three nanotubes and azimuthally
averaging the result.

The electrostatic portion of the potential in the ICs contains two
contributions, denoted $U_1$ and $U_2$, which arise, respectively, from
the interaction of the H$_2$ quadrupole moment with the local
electrostatic field of the SWNT, and the interaction of the
H$_2$ static multipole moments with the screening charges induced on
the surface.  $U_1$ is given by

\beq
 U_1( z_c, \cos\theta ) = - \frac{1}{6}\sum_{i , j} \Theta_{ij}
 \frac{\partial^2 \Phi(z_c)}{\partial x_i \partial x_j} \, ,
\eeq

\noindent in which the $\Theta_{ij}$ are the components of the
quadrupole moment and $\Phi$ is the local electrostatic field, which
was calculated from first principles for ``zigzag'' SWNT with radius
$R=6.9$~\AA\cite{drag}.  To evaluate $U_2$, we included the H$_2$
quadrupole moment (which is the first nonzero permanent moment of an
axial quadrupole), and the hexadecapole moment. This problem is quite
complicated, so we introduce three simplifications.  Due to the small
size of the H$_2$ molecule, we consider its multipoles to be
``point"-like.  Next, to simplify the complex geometry of the IC, we
represent the SWNT walls with graphene planes.  The locations of the
image charges are determined by the boundary condition requiring that
the potential be zero at each plane.  Finally, we assumed that the
graphene planes are perfectly conducting, so that the static and
dynamic responses are represented by images.  We adopted the same value
for $z_0$ and the same form for the interaction as in the previous
case.

To compute the vibrational frequency shift, the total holding potential
$U$ must be expressed as a function of the relative bondlength change,
$\xi=(r-r_e)/r_e$, relative to the equilibrium value, $r_e$, and also
of the configuration $\tau$ of the molecule.  Given a sufficiently
small H$_2$-substrate interaction, one can treat $U$ and the anharmonic
terms in the potential energy function of the free H$_2$ molecule as
perturbations to the harmonic oscillator Hamiltonian\cite{buck}. Using
first- and second-order perturbation theory, the change in the
frequency of the fundamental (n=0$\rightarrow$n=1) transition due to
$U$ is\cite{buck,herm}:

\beq
\label{buckinghameqn}
\Delta\omega = \Delta\omega_{1\leftarrow0} = 
	\frac{B_e}{\hbar\omega_e}<U''-3 a U'>_\tau + 
	O(\frac{B_e}{\hbar\omega_e})^2 \, .
\eeq

\noindent in which $B_e$ is the equilibrium rotational constant, $a$ is
the anharmonicity, primes denote derivatives with respect to $\xi$, and
the angled brackets denote the average over all $\tau$.  The
expressions for $\Theta(r)$ and $\Phi(r)$ were derived via a model in
which the electron cloud is concentrated between the nuclei rather than
about each individual nucleus.  Assuming an H$_2$ bondlength of
$0.71$~\AA, and placing the electrons symmetrically on the molecular
axis, separated by $0.48$~\AA, the experimental value of the H$_2$
quadrupole moment is reproduced.  From this model, we find $\Theta\sim
r^2$ and $\Phi\sim r^4$, for small changes in $\xi$ about $r_e$.

Regarding the orientational dependence, $U'$ and $U''$ are functions of
the $z$-displacement of molecular center-of-mass, and of the polar
($\theta$) and azimuthal ($\phi$) angles.  $U$ depends on $z_c$, so the
translational and rotational parts of $U'(z_c;\theta,\phi)$,
$U''(z_c;\theta,\phi)$ do not decouple exactly.  To separate the
variables, we treat the interaction of translational and rotational
modes in a mean-field manner\cite{nov}.  The effective interaction was
computed by averaging the derivatives of Eqn.\ref{buckinghameqn} over
the center-of-mass vibrations of the molecule.  Treating these as modes
of a simple harmonic oscillator, the  $rms$ deviations in the
$z$-position are found to be $\delta z_{rms} = 0.298$~\AA~ on a single
graphene surface, and $\delta z_{rms}=0.225$~\AA~ for H$_2$ in an IC.
Next, to compute $<U'>_{\tau}$ and $<U''>_{\tau}$, variations were
performed with respect to $r$ in the adsorption potential for different
$\tau$.  Finally,  we averaged over all orientations ($\theta$,$\phi$)
of the molecular axis.

Using the values a=-1.6 and $B_e/\hbar\omega_e$=0.0138\cite{herm}, we
find that for the case of H$_2$ adsorbed on graphene, our procedure
predicts an upshift of $\Delta\omega_{1\leftarrow0}$=1.4~cm$^{-1}$
($<3aU'>=-126.2$~cm$^{-1}$, $<U''>=-28.3$~cm$^{-1}$).  In contrast, for
H$_2$ adsorbed in an IC, we predict a {\it downshift} of
$\Delta\omega_{1\leftarrow0}$=-2.9~cm$^{-1}$ ($<3aU'>=168.0$~cm$^{-1}$,
$<U''>=-42.2$~cm$^{-1}$).  We expect similar results for D$_2$:  the
same potential parameters and quadrupole moment may be assumed, though
$a$ and $B_e/\hbar\omega_e$ are smaller\cite{herzberg}, and should lead
to smaller shifts.  In the case of HD, the center of mass and charge do
not coincide, leading to a larger Lennard-Jones contribution to the
potential and a larger upshift.

%----------------------------------------------------------CONCLUSIONS

In summary, we have observed several lines within the rotational and
vibrational Raman spectra of H$_2$, HD, and D$_2$ adsorbed on SWNT;
observed frequency shifts are small and consistent with physisorption.
Comparative Q-branch data for H$_2$ on C$_{60}$ reveal substructure in
the Q-branch, which appears to arise from H$_2$ adsorbed in different
sites. An interaction potential was developed to estimate the frequency
shifts of the adsorbate vibrational modes; our results support the
multiple-site interpretation of our data.
We wish to thank Vin Crespi, Roger Herman, John Lewis, and Dragan
Stojkovic for their generous contributions to this work.  This work was
funded by the Department of Energy and the Army Research Office.  PCE
and KAW were funded, in part, by the Office of Naval Research.

%---------------------------------------------------------------REFS

\newpage
%----------------------------------------------------------------Tables
\begin{table}[t]
\begin{center}
\begin{tabular}{lcccc}
Adsorbate & Transition		& $\nu_{free}$ 		& $\nu_{ads}$ 		& $\Delta\nu$ 		\\
species	  & 			& (cm$^{-1}$)		& (cm$^{-1}$)		& (cm$^{-1}$)		\\
\hline\\
H$_2$	  & Q(0)		& 4160.5		& 4162.3		& +1.8			\\
 	  & 			&  			& 4161.1		& +0.6			\\
 	  & 			& 			& 4159.3		& -1.2			\\
	  & Q(1)		& 4154.4		& 4156.4		& +2.0			\\
	  &			&			& 4155.2		& +0.8			\\
	  &			&			& 4153.5		& -0.9			\\
	  & S(0)		& 354.4			& 354.8			& +0.4			\\
	  & S(1)		& 587.0			& 587.5			& +0.5			\\
\hline\
\\
HD	  & Q(0)		& 3629.8		& 3632.5		& +2.7			\\
 	  & Q(1)		& 3625.8		& 3628.8		& +0.6			\\
 	  & Q(2)		& 3620.5		& 3620.9		& +0.4			\\
	  & S(0)		& 267.1			& 267.7			& +0.6			\\
 	  & S(1)		& 443.1			& 444.1			& +1.0			\\
\hline\
\\
D$_2$	  & Q(0)		& 2993.5 		& 2993.8		& +0.3 		\\
 	  & Q(1)		& 2991.4 		& 2991.9		& +0.5 			\\
	  & Q(2)		& 2987.2		& 2987.5		& +0.3			\\
	  & S(0)		& 179.1			& 178.5			& -0.6			\\
 	  & S(1)		& 297.5			& 297.8			& +0.3			\\
\hline\
\end{tabular}
\end{center}
\end{table}

%----------------------------------------------------------------FIGURES

\begin{figure}
\label{fig1}
\centerline{
\epsfxsize=3.50 in
\epsfbox{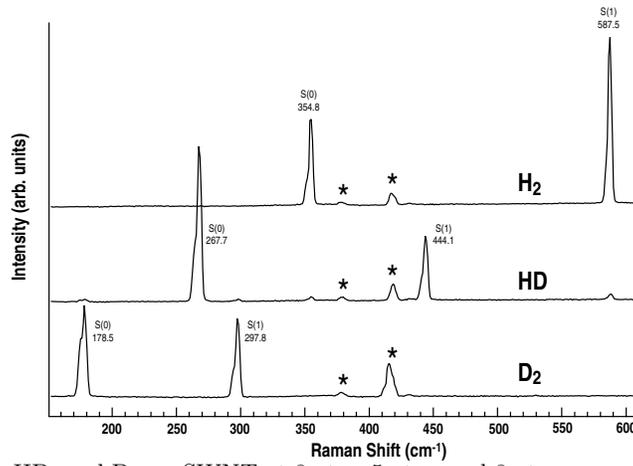}}
\caption{S-branch data for H$_2$, HD, and D$_2$ on SWNT at 8~atm, 5~atm, 
and 8~atm, respectively.  Asterisks indicate unrelated lines due to external 
calibration sources.}
\end{figure}

\begin{figure}
\label{fig2}
\centerline{
\epsfxsize=3.50 in
\epsfbox{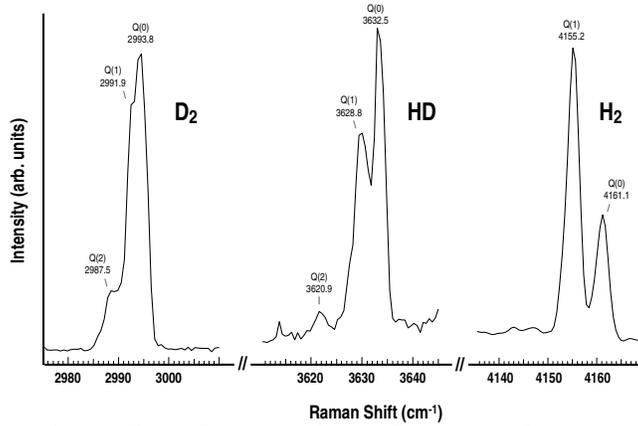}}
\caption{Q-branch data for H$_2$, HD, and D$_2$ on SWNT at 8~atm, 5~atm, 
and 8~atm, respectively.  For comparisons of these peak positions with
the frequencies of the free molecules, see Table~1.}
\end{figure}

\begin{figure}
\label{fig3}
\centerline{
\epsfxsize=3.50 in
\epsfbox{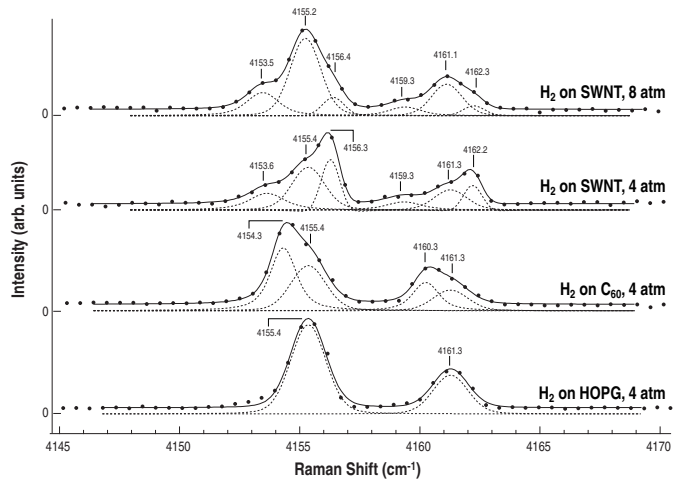}}
\caption{Q-branch data for H$_2$ on SWNT at 8 and 4~atm, together with
data for C$_{60}$ and HOPG at 4~atm.}
\end{figure} 

\end{document}